\documentclass[a4paper,onecolumn,nofootinbib,preprintnumbers]{revtex4}

\usepackage[top=2.8cm, bottom=2.8cm, left=2.4cm, right=2.4cm]{geometry}
\usepackage[utf8]{inputenc}  
\usepackage[T1]{fontenc}     
\usepackage{amsmath,amssymb,lmodern} 
\usepackage{graphicx}
\usepackage{hyperref}
\usepackage{color}

\newcommand{\ba}{\begin{eqnarray}}
\newcommand{\ea}{\end{eqnarray}}
\newcommand{\be}{\begin{equation}}
\newcommand{\ee}{\end{equation}}
\newcommand{\bea}{\begin{eqnarray}}
\newcommand{\eea}{\end{eqnarray}}
\newcommand{\beq}{\begin{equation}}
\newcommand{\eeq}{\end{equation}}
\newcommand{\beqar}{\begin{eqnarray}}
\newcommand{\eeqar}{\end{eqnarray}}
\newcommand{\beqars}{\begin{eqnarray*}}
\newcommand{\eeqars}{\end{eqnarray*}}
\newcommand{\bc}{\begin{center}}
\newcommand{\ec}{\end{center}}
\newcommand{\ben}{\begin{enumerate}}
\newcommand{\een}{\end{enumerate}}
\newcommand{\bit}{\begin{itemize}}
\newcommand{\eit}{\end{itemize}}
\newcommand{\bw}{\begin{widetext}}
\newcommand{\ew}{\end{widetext}}
\newcommand{\bcl}{\begin{columns}}
\newcommand{\ecl}{\end{columns}}

\newcommand{\dd}{\mbox{d}}

\newcommand{\A}{{\cal A}}

\begin{document}

\preprint{PI/UAN-2015-589FT}

\title{Generalized Proca action for an Abelian vector field}

\author{Erwan Allys}
\email{allys@iap.fr}
\author{Patrick Peter}
\email{peter@iap.fr}
\affiliation{Institut d'Astrophysique de Paris, UMR 7095, 98 bis bd
  Arago, 75014 Paris, France \\ UPMC Universit\'e Paris 6 et CNRS,
  Sorbonne Universit\'es}

\author{Yeinzon Rodr\'iguez}
\email{yeinzon.rodriguez@uan.edu.co}
\affiliation{Centro de Investigaciones en Ciencias B\'asicas y Aplicadas, Universidad Antonio Nari\~no, \\ Cra 3 Este \# 47A-15, Bogot\'a D.C. 110231, Colombia}
\affiliation{Escuela  de  F\'isica,  Universidad  Industrial  de  Santander, \\ Ciudad  Universitaria,  Bucaramanga  680002,  Colombia}
\affiliation{Simons Associate at The Abdus Salam International Centre for Theoretical Physics, \\ Strada Costiera 11, I-34151, Trieste, Italy}

\date{\today}

\begin{abstract}
We revisit the most general theory for a massive vector field with
derivative self-interactions, extending previous works on the subject
to account for terms having trivial total derivative interactions
for the longitudinal mode.
In the flat spacetime (Minkowski) case, we obtain all the possible terms
containing products of up to five first-order derivatives of the vector
field, and provide a conjecture about higher-order terms. Rendering the
metric dynamical, we covariantize the results and add all possible terms
implying curvature.
\end{abstract}

\maketitle
\section{Introduction}

Modifying general relativity in order to account for otherwise unresolved issues like,
e.g., the cosmological constant or even dark matter, has become common
practice lately. One way of doing so, following the scalar-tensor
proposal of the \oldstylenums{1930}s, is the so-called galileon method,
which provides a means to write down the most general theory allowing
for Ostrogradski-instability-free \cite{ostro,Woodard:2006nt}
second-order equations of motion
\cite{Nicolis:2008in,Deffayet:2009wt,Deffayet:2009mn,Deffayet:2011gz,Deffayet:2013lga},
 which happens to be equivalent \cite{Kobayashi:2011nu}, in the
 single-field case\footnote{It is a subset in the multi-field case.} and
 in four dimensions, to that proposed much earlier by Horndeski
 \cite{Horndeski:1974wa}.  The latter author, indeed, had generalized
 this idea to what, today, should appropriately be named ``vector
 Galileon'', an Abelian vector field, with an action including sources,
 and with the assumption of recovering the Maxwell equations in flat
 spacetime \cite{Horndeski:1976gi}.  Related works were developed
 recently that extended the applicability of the word ``vector
 Galileon'':  not invoking gauge invariances but having several vector
 fields and sticking to purely second-order field equations
 \cite{Deffayet:2010zh}, coupling an Abelian vector field with a scalar
 field in the framework of Einstein gravity \cite{Fleury:2014qfa}, or
 invoking an Abelian gauge invariance for just one vector field in flat
 spacetime and sticking to purely second-order field equations
 \cite{Deffayet:2013tca}.  The latter work became a no-go theorem that,
 however, is not
the end of the vector Galileon theories:
it suffices to drop the U(1) invariance hypothesis to obtain more terms
in a non-trivial and viable theory. This procedure thus naturally
generalizes the Proca theory for a massive vector field in the sense
that it describes a vector field with second-order equations of motion
for three propagating degrees of freedom, as required for the finite-dimensional representation of the Lorentz group leading to a massive
spin 1 field.

In previous works \cite{Heisenberg:2014rta,Tasinato:2014eka}, such a
theory was elaborated, introducing some fruitful ideas. A thorough
examination of these works revealed that some terms that have been
proposed are somehow redundant, while others appear to be missing. The
purpose of the present article is thus to complete these pioneering
works, investigating in a hopefully exhaustive way all the terms which
can be included in such a generalized Proca theory under our
assumptions\footnote{Some general works on general ghost-free Lagrangians 
published recently, see Refs. \cite{Li:2015vwa,Li:2015fxa}, can also be applied to the generalization of the Proca theory.}. The most crucial constraint in the building up of the
theory is the demand that only three polarizations should be able to
propagate, namely two transverse and one longitudinal, the latter
being reducible to the scalar formulation of the Galileon, thus providing
the required link between the vector and scalar theories.
On the other hand, considering the transverse modes, we found
numerous differences between the two theories, as in particular we
obtained many more terms in the vector case. Indeed, whereas the scalar
Lagrangian comprises a finite number of possibilities, we conjecture
that the vector one is made up of an infinite tower of such terms.

The structure of the rest of the article goes as follows. First, we recall
previous works by introducing the generalized Abelian Proca theory,
and explicit the assumptions under which our Lagrangian is constructed.
Recalling as a starting point the terms already obtained in Refs. \cite{Heisenberg:2014rta}
and \cite{Tasinato:2014eka}, we use their results to understand
what kind of other terms could also be compatible with our assumptions.
We then move on to these extra terms in Sec. \ref{AdditionalTerms}, first
introducing a general method of investigation of these new terms, then
applying the latter to terms containing products of up to five first-order
derivatives of the vector field. We conclude this part by discussing what 
the complete generalized Proca theory could be. Finally, in
Sec. \ref{Covariantization}, we consider the extension of the Proca
theory to a curved spacetime. We need to add yet more terms necessary
to render the Lagrangian healthy (in the sense defined above), and this
also leads to new interactions with gravity.

\section{Generalized Abelian Proca theory}

The Proca theory describes the dynamics of a vector field with a mass
term in its Lagrangian, thus explicitly breaking the U(1) gauge
invariance usually associated with the Maxwell field $A_\mu$. The Proca
action therefore reads
\begin{equation}
\mathcal{S}_\mathrm{Proca}  = \int \mathcal{L}_\mathrm{Proca}\,\dd^4x 
= \int \left( -\frac14 F_{\mu\nu} F^{\mu\nu} + \frac12 m_A^2 X \right)\,\dd^4x,
\label{Proca}
\end{equation}
where the antisymmetric Faraday tensor is $F_{\mu\nu} \equiv
\partial_\mu A_\nu -\partial_\nu A_\mu$, and we have defined, for further
convenience, the shorthand notation $X\equiv A_\mu A^\mu$. Following and
extending Ref.~\cite{Heisenberg:2014rta}, we want to provide the
broadest possible generalization of this action under a set of
specific assumptions to which we now turn.

Before we move on to curved spacetime in Sec.~\ref{Covariantization},
we shall in what follows consider that our field lives in a non-dynamical
Minkowski metric $g^{\mu \nu}=\eta^{\mu \nu}=\rm{diag}(-1,+1,+1,+1)$. We shall
also make use of the notation $(\partial \cdot A)=\partial_\mu A^\mu$
for simplification and notational convenience.

\subsection{Theoretical assumptions}
\label{conditions}

The idea is to add to the minimal Proca action in Eq. (\ref{Proca}) all
acceptable terms containing not only functions of $X$ but also
derivative self-interactions under a set of suitable conditions.
In order to explicit those conditions, let us split the vector into
a scalar/vector decomposition
\begin{equation}
A_\mu=\partial_\mu \pi + \bar{A}_\mu \ \ \ \hbox{or} \ \ \ A
= \dd \pi + \bar{A},
\end{equation}
 where $\pi$ is a scalar field, called the St\"uckelberg field, and
 $\bar{A}_\mu$ is a divergence-free vector ($\partial_\mu \bar
 A^\mu=0$), containing the curl part of the field, i.e. that for which
 the Faraday tensor is non vanishing (in differential form terms, it is
 the non-exact part of the form, with the Faraday form now equal to
 $F=\dd A = \dd\bar{A}$). The conditions we want to impose on the theory
 in order that it makes (classical) sense are reminiscent of the
 galileon conditions, namely we demand
\begin{itemize}
\item[a)] at most second-order equations of motion for all physical
degrees of freedom, i.e., for both $A_\mu$ and $\pi$,
\item[b)] at most second-order derivative terms in $\pi$ in the action,
and first-order derivative terms for $A_\mu$,
\item[c)] only three propagating degrees of freedom for the vector field;
 in other words, there should be no propagation of its zeroth component.
\end{itemize} 
The first condition ensures stability, as discussed e.g. in
Ref.~\cite{Woodard:2006nt}\footnote{This result is indirect: accepting a
third-order derivative in the equations of motion would imply at least a
second-order time derivative in the Lagrangian, which is not degenerate
but yields a Hamiltonian which is unbounded from below.}. Because we
demand first-order derivatives in $A_\mu$, the second condition implies
that the first is automatically satisfied for $A_\mu$; the first
condition is therefore necessary to implement only for the scalar part
$\pi$ of the vector field. As for the third condition, it stems from the
definition of a vector as a unit spin Lorentz-group representation, a spin
$s$ object having $(2s+1)$ propagating degrees of freedom. In Ref.
\cite{Heisenberg:2014rta}, an extra condition was implicit, namely that the
longitudinal mode should not have trivial total derivative interactions; the
terms we obtain below and that were not written down in this reference
stem precisely from our relaxing of this condition.

For a given Lagrangian function $\mathcal{L}(A_\mu)$, the number of
actually propagating degrees of freedom must be limited to three. To enforce
this requirement, we compute the Hessian matrix $\mathcal{H}^{\mu\nu}$
associated with the Lagrangian term considered through
\begin{equation}
\mathcal{H}^{\mu\nu}=\frac{\partial^2 \mathcal{L}}{\partial(\partial_0
A_\mu)\partial(\partial_0A_\nu)},
\label{Hess}
\end{equation}
and demand that it should have a vanishing eigenvalue associated to the
absence of propagation of the time component of $A_\mu$. A sufficient
condition to achieve this goal is to ask that $\mathcal{H}^{00}=0$ and
$\mathcal{H}^{0i}=0$. One can show that this condition is the unique one
permitting a vanishing eigenvalue: since different derivatives of
the vector field are independent, the only way to diagonalize the
Hessian without any vanishing eigenvalue is to cancel its symmetric terms
with one another, and this, in turn, is not possible by means of a Lorentz
transformation.

\subsection{The Heisenberg action}

Let us summarize Ref.~\cite{Heisenberg:2014rta}, switching to notations
agreeing with those we use here. The Lagrangian terms which appear in
addition to the standard kinetic term will be called $\mathcal{L}_n$, a
notation coming from that which is usual in the Galileon theory
\cite{Deffayet:2013lga}. There, $N = n + 2$ is the number of scalar
field factors appearing in each Lagrangian, and $M=n-2$ is the number of
second derivatives of the Galileon field (we do not count the
arbitrary functions of $X$ in this power counting). Keeping the same
notation,
now applied to the scalar part $\pi$ of the vector field, one can write
the general Lagrangian as the usual Maxwell Lagrangian with a
generalized mass term, i.e., one
replaces the Lagrangian in Eq.~(\ref{Proca}) by
\begin{equation}
\mathcal{L}_\mathrm{Proca}^\mathrm{gen.}=
-\frac{1}{4}F_{\mu\nu}F^{\mu\nu}+\sum_{n=2}^5
\mathcal{L}_n,
\end{equation}
with
\begin{equation}
\begin{split}
\mathcal{L}_2 = & f_2(A_\mu, F_{\mu\nu},\tilde{F}_{\mu\nu}),\cr
\mathcal{L}_3 = & f_3^{\mathrm{Gal}}(X) (\partial \cdot A),\cr
\mathcal{L}_4 = & f_4^{\mathrm{Gal}}(X) \left[ \left( \partial_\mu A_\nu
                                 \partial^\nu A^\mu \right) -
                                 \left( \partial\cdot A \right) \left( \partial\cdot A \right) \right],\cr                                 
\mathcal{L}_5 = & f_5^{\mathrm{Gal}}(X) \left[\left(\partial\cdot A\right)^3 - 3 \left(\partial\cdot A\right)
  \left(\partial_ {\nu}A_ {\rho} \partial^ {\rho}A^ {\nu}\right) + 2
  \left(\partial_{\mu}A^ {\nu} \partial_ {\nu}A^ {\rho} \partial_
       {\rho}A_ {\mu}\right) \right] \cr  &+ f_5^{\text{Perm}}(X)
\left[ \left(\partial\cdot A\right) \left[ \left(\partial_ {\rho}A_
    {\nu} \partial^ {\rho}A^ {\nu}\right) - \left(\partial_ {\nu}A_
    {\rho} \partial^ {\rho}A^ {\nu}\right) \right]+ \left(\partial_
  {\mu}A_ {\rho} \partial^ {\nu}A^ {\mu} \partial^ {\rho}A_ {\nu}
  \right) - \left(\partial^ {\nu}A^ {\mu} \partial_ {\rho}A_ {\nu}
  \partial^ {\rho}A_ {\mu}\right)\right],
\end{split}
\label{LagHeisenberg}
\end{equation}
where $f_2(A_\mu, F_{\mu\nu},\tilde{F}_{\mu\nu})$ is an arbitrary
function of all scalars which can be constructed from $A_\mu$,
$F_{\mu\nu}$  and $\tilde{F}_{\mu\nu}\equiv
\frac{1}{2}\epsilon_{\mu\nu\alpha\beta}F^{\alpha\beta}$, the latter
being the Hodge dual of the Faraday tensor\footnote{Here and in
what follows, we denote by $\epsilon^{\sigma_1\sigma_2\dots \sigma_D}$
the totally antisymmetric Levi-Civita tensor in $D$ dimensions (in the
case $D=4$ we are concerned with in the rest of this paper, we write
$\epsilon^{\mu\nu\alpha\beta}$).}. All other functions $f$ are
independent arbitrary functions of $X$ only.

These expressions actually differ from those given in
Ref.~\cite{Heisenberg:2014rta}. First, we did not add a term of the form
$f(X) \left[ \left(\partial_ {\nu}A_ {\mu} \partial^ {\nu}A^ {\mu}\right) -
\left(\partial_ {\mu}A_ {\nu} \partial^ {\nu}A^ {\mu}\right) \right]$ in
$\mathcal{L}_4$, as such a term is merely equal to
$\frac12 f(X) F_{\mu\nu}F^{\mu\nu}$ and is therefore
already contained in $f_2$. Moreover, both terms in $\mathcal{L}_5$
have different arbitrary functions as prefactors. In fact, this general
form is sufficient to verify the conditions given in Sec.
\ref{conditions}; this will be explained in Sec. \ref{L3->5}.

The Lagrangians given by Eqs. (\ref{LagHeisenberg}) contain different
kinds of terms with various origins. The first contributors, with
prefactors given by the arbitrary functions $f_n^{\text{Gal}}$,
originate from the scalar part of the vector field:  setting $\bar{A}\to
0$, i.e. $A_\mu\to \partial_\mu \pi$, they are indeed nonvanishing. As,
for consistency, they must verify the hypothesis of Galileon theory,
they should thus come straightforwardly from the Galileon theory and
therefore ought to be equivalent to the only Lagrangians satisfying such
properties as they have been classified in Ref.~\cite{Deffayet:2011gz}.
One way of defining them explicitly consists, for instance, in taking the
specific form chosen e.g. in Ref.~\cite{Deffayet:2011gz,Deffayet:2013lga} (where it was
called $\mathcal{L}^3_N$)
\begin{equation}
\label{LagGalileon}
\mathcal{L}^\text{Gal}_{N} =\left[ A_{(2n)}^{\mu_1\mu_2\dots
  \mu_n\nu_1\nu_2\dots \nu_n} \pi_{\lambda} \pi^{\lambda}
\right] \pi_{\mu_1 \nu_1} \dots \pi_{\mu_n \nu_n},
\end{equation}
with 
\begin{equation}
A_{(2m)}^{\mu_1\mu_2\dots \mu_m\nu_1\nu_2\dots \nu_m} =
\frac{1}{(D-m)!} \epsilon^{\mu_1\mu_2\dots \mu_m\sigma_1\sigma_2\dots
  \sigma_{D-m}} \epsilon^{\nu_1\nu_2\dots \nu_m}{}_{ \sigma_1\dots
  \sigma_{D-m}},
\end{equation}
where $\pi_{\mu\nu\cdots}\equiv\partial_\mu\partial_\nu\cdots\pi$ is
a convenient shorthand notation for the derivatives. 
The definition which appears to be unambiguous then arises from the
following procedure: we substitute, in Eq.~\eqref{LagGalileon}, the
first derivative of the scalar field by a component of the vector field
($\pi_{\mu\nu\cdots}\to A_{\mu,\nu\cdots}$), keeping all indices the way
they appear in the original form. From the point of view of the scalar
part of the Lagrangian, this leads to the same Galileon action, and it
would also if we were to permute derivatives (indices in this case). For
example, we could take in $\mathcal{L}_5$, a term having the special
form
\begin{equation}
f_5^{\text{Gal,alt}}(X) \left[\left(\partial\cdot A\right)^3 -
  \left(\partial\cdot A\right) \left[ \left(\partial_ {\nu}A_ {\rho}
    \partial^ {\rho}A^ {\nu}\right) + 2 \left(\partial_ {\rho}A_ {\nu}
    \partial^ {\rho}A^ {\nu}\right)\right] + 2 \left(\partial^ {\nu}A^
       {\mu} \partial_ {\rho}A_ {\nu} \partial^ {\rho}A_ {\mu}\right)
       \right] ,
\end{equation}
which would also reduce to Eq.~(\ref{LagGalileon}) in the scalar case.
This term is also obtained by taking
$f_5^{\text{Perm}}=-2f_5^{\text{Gal}}$. Both terms contained in
$\mathcal{L}_5$ could thus be obtained by using the Galileon Lagrangian
and shuffling around the second-order derivatives before doing the
replacement by the vector field. The choice proposed above however provides a
uniquely defined action when it comes to the vector field case.

The second category of terms, including $f_2$ (apart from terms containing
only $X$) and $f_5^{\text{Perm}}$, gives a vanishing
contribution when going to the scalar sector, due to the fact that $
\partial_\mu \partial_\nu \pi = \partial_\nu \partial_\mu \pi$, which is
not verified by $\partial_\mu \bar{A}_\nu$ by definition. This is
trivial for the terms containing the Faraday tensor, since $A_\mu\to
\partial_\mu\pi$ implies $F_{\mu\nu}\to 0$ identically, and can be
verified explicitly for the term including $f_5^{\text{Perm}}$. These
terms can also be seen to lead to only three propagating degrees of
freedom for the vector field. Given this fact, one is naturally led to
ask whether similar kind of terms, including more derivatives, could be
possible. They are discussed in the following section.

\section{Additional terms}
\label{AdditionalTerms}
\subsection{Procedure of investigation}
\label{procedure}

In order to find all the possible terms satisfying the conditions of
Sec. \ref{conditions}, we developed a systematic procedure, which we
explain below by considering the term $\mathcal{L}_n$ of the
Lagrangian, containing $(n-2)$ first-order derivatives of the vector
field\footnote{The index $n$ appearing in our Lagrangians is inconvenient
but we keep it in order to follow previous conventions developed historically
from the Galileon action.}.
The final action will then consist of a sum over all such possible terms,
weighted by arbitrary functions of $X$, as this can in no way change
anything in the discussion of the Hessian.

We first list all the possible terms which can be written as
contractions of $(n-2)$ first-order derivatives of $A_\mu$. We call them
$\mathcal{L}^{\text{test}}_{n,i}$, where $i$ labels the different terms
which can be written for a given $n$. These test Lagrangians are then
linearly combined to provide the most general term at a given order $n$,
whose Hessian, in Eq. \eqref{Hess}, is computed. We apply our requirements
($\mathcal{H}^{0\mu}=0$ for all $\mu=0,\cdots,3$) to derive relations
among the coefficients of the linear combination, and to finally obtain
the relevant terms giving only three propagating degrees of freedom.
Those fall into two distinct categories, namely those whose scalar
sector vanishes or not. The latter necessarily reduce to the
Galileon action, while the former are new, purely vectorial, terms
which, in turn, can either be constructed from $F_{\mu\nu}$ only, or be
the terms we are interested in. Any term leading to a non-trivial
dynamics for the scalar part that would be nonvanishing should be then
set to zero in order to comply with the requirement that the scalar
action is that provided by the Galileon.

We are investigating terms that contain only derivatives  and not the
vector field itself in the contractions over Lorentz indices, as all
possible such terms can always be reduced to those studied below and a
total derivative. The typical example we have in mind is a term of the
form
$$
A_\alpha A^\mu\partial_\mu A^\alpha = \frac12 \left[ \partial_\mu
\left( X A^\mu\right) - X \left( \partial \cdot A\right) \right],
$$
which is then equivalent to those we discuss below, up to a total
derivative. Note that a similar term including an arbitrary function of
$X$ in front would lead to second-order derivatives in $A$, and is thus
excluded by construction.

Finally, we would like to mention at this point that we are here only
considering terms involving the metric, i.e., we build scalars for the
Lagrangian by contracting the indices of quantities such as
$\partial_\mu A_\nu$ using the metric $g^{\mu\nu}$. Below, in
Sec.~\ref{sec:eps}, we will consider another possible case, that for
which scalars are also built through contractions with the completely
antisymmetric tensor $\epsilon^{\alpha\beta\mu\nu}$; these extra
terms and those first considered produce independent Hessian variations
and can thus be studied independently.

\subsection{From $\mathcal{L}_3$ to $\mathcal{L}_5$}
\label{L3->5}

The first term appearing in our expansion is $\mathcal{L}_3$, whose only
test Lagrangian can be written as $(\partial \cdot A)$, which has a
vanishing Hessian, and is nothing but the Galileon vector term
\begin{equation}
\mathcal{L}_{3}^{\text{Gal}} =  \left(\partial \cdot A\right).
\end{equation}

The case of $\mathcal{L}_4$ is slightly more involved, the test Lagrangians
reading
\begin{equation}
\mathcal{L}^{\text{test}}_{4,1} = \left(\partial\cdot A\right)^2,\ \ \ \ 
\mathcal{L}^{\text{test}}_{4,2} = \left(\partial_\nu A_\mu \partial^\nu A^\mu \right),\ \ \ \ \hbox{and}
\ \ \ \ \mathcal{L}^{\text{test}}_{4,3} = \left(\partial_\mu A_\nu \partial^\nu A^\mu \right).
\end{equation}
Setting the full test Lagrangian at this order to be
\begin{equation}
\mathcal{L}_{4}^{\text{test}} = x_1 \mathcal{L}^{\text{test}}_{4,1} + x_2
\mathcal{L}^{\text{test}}_{4,2} + x_3 \mathcal{L}^{\text{test}}_{4,3},
\end{equation}
we find the following relevant Hessian terms,
\begin{equation}
\mathcal{H}_{4}^{00} = 2 \left(x_{1} + x_{2} + x_{3}\right) \ \ \ \ \hbox{and}
\ \ \ \ \mathcal{H}_{4}^{0i} = 0,
\end{equation}
so that the solutions ensuring only three propagating degrees of freedom are
\begin{equation}
\mathcal{L}_{4}^{\text{Gal}} = \left(\partial_ {\mu}A_ {\nu} \partial^
        {\nu}A^ {\mu}\right) - \left(\partial\cdot A\right)
        \left(\partial\cdot A\right) ,
\end{equation}
and
\begin{equation}
\mathcal{L}_{FF} = F_{\mu\nu}F^{\mu\nu}=\left(\partial_ {\nu}A_ {\mu}
\partial^ {\nu}A^ {\mu}\right) - \left(\partial_ {\mu}A_ {\nu}
\partial^ {\nu}A^ {\mu}\right).
\end{equation}
We identify the first term with the Galileon vector term, and the second
to the only term at this order which can be built from the field tensor.
This term was expected, and we will
not consider it any further since it is already included in
$\mathcal{L}_{2}$. We also check that, as these terms independently
verify the required conditions, they are effectively independent.

Let us move on to the following order, namely $\mathcal{L}_5$. The test
Lagrangians are
\begin{equation}
\begin{split}
\mathcal{L}^{\text{test}}_{5,1} & = \left(\partial\cdot A\right)^3,\\
\mathcal{L}^{\text{test}}_{5,2} & = \left(\partial\cdot A\right) \left(\partial_
       {\rho}A_ {\nu} \partial^ {\rho}A^\nu\right),\\
\mathcal{L}^{\text{test}}_{5,3} & = \left(\partial\cdot
        A\right) \left(\partial_ {\nu}A_ {\rho} \partial^ {\rho}A^\nu\right),\\
\mathcal{L}^{\text{test}}_{5,4} & = \left(\partial_ {\mu}A_
        {\rho} \partial^ {\nu}A^ {\mu} \partial^ {\rho}A_\nu\right),\\
\mathcal{L}^{\text{test}}_{5,5} & = \left(\partial^ {\nu}A^
        {\mu} \partial_ {\rho}A_ {\nu} \partial^ {\rho}A_\mu\right).
\end{split}
\end{equation}
Using 
\begin{equation}
\mathcal{L}_{5}^{\text{test}} = x_{1} \mathcal{L}^{\text{test}}_{5,1} +
x_{2} \mathcal{L}^{\text{test}}_{5,2} + x_{3}
\mathcal{L}^{\text{test}}_{5,3} + x_{4} \mathcal{L}^{\text{test}}_{5,4}
+x_{5} \mathcal{L}^{\text{test}}_{5,5},
\end{equation}
we obtain the relevant Hessian terms\footnote{Although the Hessian looks
different from that of Ref.~\cite{Heisenberg:2014rta}, it actually is
equivalent, the difference stemming from the fact that in the latter work,
$\mathcal{H}_{5}^{00}$ was decomposed on $\partial_i A^i$ and
$\partial_0 A^0$, while we chose to decompose it on $\partial_\mu A^\mu$
and $\partial_0 A^0$.}
\begin{equation}
\begin{split}
\mathcal{H}_{5}^{00} & = -2 \left(-3 x_{1} - x_{2} - x_{3}\right)
\left(\partial \cdot A\right) -2 \bigl[2 x_{2} + 2 x_{3} + 3
  \left(x_{4} + x_{5}\right) \bigr]
\left(\partial^{0}A^{0}\right),\\
\mathcal{H}_{5}^{0i} & = -2
\left(x_{2} + x_{5}\right) \left(\partial^{0}A^{i}\right) - \left(2
x_{3} + 3 x_{4} + x_{5}\right) \left(\partial^{i}A^{0} \right),
\end{split}
\end{equation}
which, when vanishing, provide the following solutions:
\begin{equation}
\mathcal{L}_{5}^{\text{Gal}} = \left(\partial\cdot A\right)^3 - 3
\left(\partial\cdot A\right) \left(\partial_ {\nu}A_ {\rho} \partial^
     {\rho}A^ {\nu}\right) + 2 \left(\partial_{\mu}A^ {\nu} \partial_
     {\nu}A^ {\rho} \partial_ {\rho}A_ {\mu}\right),
     \label{Gal5}
\end{equation}
and
\begin{equation}
\mathcal{L}_{5}^{\text{Perm}} = \frac{1}{2}\left(\partial\cdot A\right) F_{\mu\nu}
F^{\mu\nu} + \partial_\rho A_\nu \partial^\nu A_\mu F^{\mu\rho}.
\label{Perm5}
\end{equation}
The first, Eq. \eqref{Gal5}, is the Galileon scalar Lagrangian, verifying
all the imposed conditions, leading to second-order equations of motion,
even multiplied by any function of $X=\partial_\mu \pi \partial^\mu
\pi$; this is shown in Ref. \cite{Deffayet:2011gz}. The second one,
Eq. \eqref{Perm5}, gives no dynamics for the Galileon scalar, since it
vanishes in the scalar sector, as can be seen from the fact that it can
be factorized in terms of the strength tensor $F_{\mu\nu}$ (but
not exclusively, since it would otherwise be gauge invariant and thus
reducible to functions of $F_{\mu\nu}F^{\mu\nu}$ and $F_{\mu\nu}\tilde
F^{\mu\nu}$). It does however also satisfy our conditions, even
multiplied by any function of $X$.

This complete the proof that Eq.~(\ref{LagHeisenberg}), as originally
obtained in Ref.~\cite{Heisenberg:2014rta}, fulfills the required constraints
for a generic classical second-order action for a vector field. We shall now
discuss the possibility of extra terms, not present in previous works.

\subsection{Higher-order actions}

Since the scalar Galileon action stops at the level discussed above, it sounds
reasonable to assume the same to apply for the vector field, in particular in
view of the fact that this field contains a scalar part, and to consider Eq. (\ref{LagHeisenberg})
to provide the most general second-order classical vector theory. This is not
what happens, in practice, as we show below, as one can indeed find extra
terms, which do vanish in the limit $\A_\mu\to\partial_\mu\pi$, leaving a
non-trivial dynamics for the divergence-free part $\bar A_\mu$.

\subsubsection{Fourth power derivatives: $\mathcal{L}_6$}
\label{L6}

We begin our examination of the higher powers in derivatives by
concentrating on $\mathcal{L}_{6}$, which involves therefore four powers
of the field gradient. Since the actual calculations imply very large
and cumbersome terms, and because those are not so important for the
understanding of this work, we have regrouped this calculation in
Appendix \ref{appL6}.

We find that there are four Lagrangians verifying the Hessian condition.
They are
\begin{equation}
\begin{split}
\mathcal{L}_{6}^{\mathrm{Gal}} =& \left(\partial\cdot A\right)^4 - 2
\left(\partial\cdot A\right) ^2 \left[\left(\partial_ {\rho}A_
  {\sigma} \partial^ {\sigma}A^ {\rho}\right) +2 \left(\partial_
  {\sigma}A_ {\rho} \partial^ {\sigma}A^ {\rho}\right) \right] + 8
\left(\partial\cdot A\right) \left(\partial^ {\rho}A^ {\nu} \partial_
     {\sigma}A_ {\rho} \partial^ {\sigma}A_ {\nu}\right) -
     \left(\partial_ {\mu}A_ {\nu} \partial^ {\nu}A^ {\mu}\right)^2 \\ &+
     4 \left(\partial_ {\nu}A_ {\mu} \partial^ {\nu}A^ {\mu}\right)
     \left(\partial_ {\rho}A_ {\sigma} \partial^ {\sigma}A^
          {\rho}\right) -2 \left(\partial_ {\nu}A^ {\sigma} \partial^
          {\nu}A^ {\mu} \partial_ {\rho}A_ {\sigma} \partial^ {\rho}A_
          {\mu}\right) - 4 \left(\partial^ {\nu}A^ {\mu} \partial^
          {\rho}A_ {\mu} \partial_ {\sigma}A_ {\rho} \partial^
          {\sigma}A_ {\nu}\right),
\end{split}
\label{Gal6}
\end{equation}
which should a priori give a dynamics for the scalar part, and
\begin{equation}
\begin{split}
\mathcal{L}_{6}^{\text{Perm}} =&  \left(\partial\cdot A\right)^2
F^{\mu\nu} F_{\mu\nu}  - \left(\partial_ {\rho}A_
     {\sigma} \partial^ {\sigma}A^ {\rho}\right)  F^{\mu\nu}F_{\mu\nu}
        + 4 \left(\partial\cdot A\right)
     \partial^\rho A^\nu \partial^\sigma A_\rho F_{\nu\sigma} \\
          & + \partial^\mu A_\nu F^\nu{}_\rho F^\rho{}_\sigma F^\sigma{}_\mu
          - 4 \,\partial^\mu A _\nu \partial^\nu A_\rho \partial^\rho A_\sigma F^\sigma{}_\mu.
\end{split}
\label{Perm6}
\end{equation}

The first term, Eq. \eqref{Gal6}, contains four second-order derivatives of
the scalar part and, thus, because it is of order higher than 3, cannot
verify the conditions of
Sec.~\ref{conditions}~\cite{Nicolis:2008in,Deffayet:2011gz}. We shall
consequently discard it.

There are two more terms which satisfy the required conditions, namely
$\mathcal{L}_{FF\cdot FF} =(F_{\mu\nu}F^{\mu\nu})^2$ and
$\mathcal{L}_{FFFF} = F^\mu{}_\nu F^\nu{}_\rho F^\rho{}_\sigma
F^\sigma{}_\mu$.  Since these terms are built out only of the
field strength tensor, they do give a dynamics to the vector field,
but are already included in $\mathcal{L}_{2}$, and we therefore will not
consider them any further.

The terms in Eq.~\eqref{Perm6} exhibit explicit powers of
($\partial \cdot A$), and therefore cannot be built
out of simple products of the field tensor. The Lagrangian
$\mathcal{L}_{6}^{\text{Perm}}$ vanishes in the scalar sector and verifies all the
conditions we want to impose, even multiplied by an arbitrary function
of $X$. We consider the general theory of a vector field, and we thus
must add to Eq. \eqref{LagHeisenberg} the term
\begin{equation}
\mathcal{L}_{6}=f_6^{\text{Perm}}(X)\mathcal{L}_{6}^{\text{Perm}},
\end{equation}
with $f_6^{\text{Perm}}$ being an arbitrary function of $X$.

\subsubsection{Fifth power derivatives: $\mathcal{L}_7$}
\label{L7}

Having found an extra non-trivial dynamical term, we now move on to
including yet one more power in the derivatives. Even more cumbersome
than those leading to $\mathcal{L}_6$, the calculations yielding
$\mathcal{L}_7$ are reproduced in Appendix \ref{appL7}. We obtain

\begin{equation}
\begin{split}
\mathcal{L}_{7}^{\text{Gal}} =& 3 \left(\partial \cdot A\right)^5-
\left(\partial \cdot A\right)^3 10\left[ \left(\partial_ {\sigma}A_
  {\gamma} \partial^ {\gamma}A^ {\sigma}\right) - 2\left(\partial_
  {\gamma}A_ {\sigma} \partial^ {\gamma}A^ {\sigma} \right)\right]+ 60
\left(\partial \cdot A\right)^2 \left(\partial^ {\sigma}A^ {\rho}
\partial_ {\gamma}A_ {\sigma} \partial^ {\gamma}A_ {\rho}\right) \\ 
& + 15 \left(\partial \cdot A\right)\left[ \left(\partial_ {\nu}A_ {\rho}
  \partial^ {\rho}A^ {\nu} \right)\left(\partial_ {\sigma}A_ {\gamma}
  \partial^ {\gamma}A^ {\sigma}\right) +2 \left(\partial_ {\rho}A_
          {\nu} \partial^ {\rho}A^ {\nu}\right)\left( \partial_
          {\sigma}A_ {\gamma} \partial^ {\sigma}A^ {\gamma}\right) -
          4\left(\partial_ {\rho}A^ {\gamma} \partial^ {\rho}A^ {\nu}
          \partial_ {\sigma}A_ {\gamma} \partial^ {\sigma}A_
                  {\nu}\right) \right. \\
& \left.  - 4\left(\partial^
                  {\rho}A^ {\nu} \partial^ {\sigma}A_ {\nu} \partial_
                  {\gamma}A_ {\sigma} \partial^ {\gamma}A_
                  {\rho}\right) \right] + 20 \left( \partial_ {\rho}A_
{\gamma} \partial^ {\sigma}A^ {\rho} \partial^ {\gamma}A_
{\sigma}\right)\left[ \left(\partial_ {\nu}A_ {\mu} \partial^ {\nu}A^
  {\mu}\right) - \left(\partial_ {\mu}A_ {\nu} \partial^ {\nu}A^
  {\mu}\right) \right] \\
& - 60 \left(\partial_ {\nu}A_ {\mu} \partial^
{\nu}A^ {\mu} \right)\left(\partial^ {\sigma}A^ {\rho} \partial_
{\gamma}A_ {\sigma} \partial^ {\gamma}A_ {\rho} \right) + 60
\left(\partial^ {\nu}A^ {\mu} \partial_ {\rho}A^ {\gamma} \partial^
     {\rho}A_ {\mu} \partial_ {\sigma}A_ {\gamma} \partial^ {\sigma}A_
     {\nu}\right)\\
& + 60 \left(\partial^ {\nu}A^ {\mu} \partial^
     {\rho}A_ {\mu} \partial^ {\sigma}A_ {\nu} \partial_ {\gamma}A_
     {\sigma} \partial^ {\gamma}A_ {\rho}\right) - 60 \left(\partial^ {\nu}
     A^{\mu} \partial_ {\rho}A_ {\gamma}
     \partial^ {\rho}A_ {\mu} \partial^ {\sigma}A_ {\nu} \partial^
             {\gamma}A_ {\sigma}\right)\\
& + 12 \left(\partial_ {\mu}A_
             {\gamma} \partial^ {\nu}A^ {\mu} \partial^ {\rho}A_ {\nu}
             \partial^ {\sigma}A_ {\rho} \partial^ {\gamma}A_
                     {\sigma}\right) ,
\end{split}
\label{Gal7}
\end{equation}
which, similarly to Eq. \eqref{Gal6}, must be discarded as it leads to higher-order
equations of motion for the scalar part, and two extra terms, namely
\begin{equation}
\begin{split}
\mathcal{L}_{7}^{\text{Perm,1}} = &  \left(\partial \cdot
A\right)^3F^{\mu\nu} F_{\mu\nu} + 6 \left(\partial \cdot A\right) ^2 
\partial^\mu A^\nu \partial^\rho A_\mu F_{\nu \rho} + 3 \left(\partial
\cdot A\right) \left[ \left(\partial_  {\nu}A_ {\rho} \partial^ {\rho}A^
{\nu} \right)^2 - \left( \partial_ {\rho}A_ {\nu} \partial^ {\rho}A^
{\nu} \right)^2 \right] \\
& + 3 \left(\partial \cdot A\right)
\left(\partial^\mu A_\nu F^\nu{}_\rho F^\rho{}_\sigma F^\sigma{}_\mu - 4
\partial^\mu A _\nu \partial^\nu A_\rho \partial^\rho A_\sigma
F^\sigma{}_\mu\right) + 4 \left(\partial_\mu A_\nu \partial^\mu A^\nu
\right)\left(\partial^\rho A^\sigma \partial_\gamma A_\rho \partial
^\gamma A_\sigma \right)\\
&  - 4 \left(\partial_\mu A_\nu \partial^\nu
A^\mu \right)\left(\partial^\rho A^\sigma \partial_\gamma A_\rho
\partial _\sigma A^\gamma \right)  + 2 \left(\partial_\mu A_\nu
\partial^\mu A^\nu \right) \left(\partial^\rho A_\sigma \partial_\gamma
A_\rho F^{\gamma \sigma} \right) \\
& - 6  \partial^\mu A_\nu
F^\nu{}_\rho F^\rho{}_\sigma F^\sigma{}_\gamma F^\gamma{}_\mu 
 + 12 \partial^\mu A_\nu \partial^\nu A_\rho \partial^\rho A_\sigma
\partial^\sigma A_\gamma F^\gamma{}_\mu ,
\end{split}
\label{1Perm7}
\end{equation}
and
\begin{equation}
\begin{split}
\mathcal{L}_{7}^{\text{Perm,2}} = & \frac{1}{4}\left(\partial \cdot
A\right)\big[\left(F_{\mu\nu} F^{\mu\nu}\right)^2 - 4\partial^\mu A_\nu
F^\nu{}_\rho F^\rho{}_{\sigma} F^\sigma{}_\mu \big] + \left(
F^{\mu\nu}F_{\mu \nu}\right) \partial^\sigma A^\rho \partial^\gamma
A_\sigma F_{\rho \gamma}  \\
&  +2 \left( \partial^\mu A_\nu
F^\nu{}_\rho F^\rho{}_\sigma F^\sigma{}_\gamma F^\gamma{}_\mu \right),
\end{split} \label{2Perm7}
\end{equation}
both of which vanish in the limit $\A_\mu\to\partial_\mu\pi$. It can
be checked explicitly that they independently verify all the conditions given in
Sec. \ref{conditions}, so the final theory will contain 
\begin{equation}
\mathcal{L}_{7}=f_{7}^{\text{Perm,1}}(X)
\mathcal{L}_{7}^{\text{Perm,1}}+f_{7}^{\text{Perm,2}}(X)
\mathcal{L}_{7}^{\text{Perm,2}},
\end{equation}
with $f_{7}^{\text{Perm,1}}$ and $f_{7}^{\text{Perm,2}}$ being two
independent arbitrary functions of $X$.

\subsection{Antisymmetric $\epsilon$ terms.}
\label{sec:eps}

As argued in Sec.~\ref{procedure}, one can also build scalars out of a
vector field by contractions with the completely antisymmetric
Levi-Civita tensor $\epsilon^{\mu\nu\alpha\beta}$. In order for these
new terms to be effectively independent on the previously derived ones,
it is necessary that two such tensors never appear contracted with one
another, since there exists relations such as $\epsilon_{\alpha\beta\mu\nu}
\epsilon^{\alpha\beta\rho\sigma} = 2 \left( \delta_\mu^\rho \delta_\nu^\sigma
-\delta_\mu^\sigma \delta_\nu^\rho \right)$, all such contractions will
give back our previous terms.
This produces only a limited number of
independent new terms at each order.

The first order, quadratic in the derivative terms, contains only one
possible contraction, namely $\mathcal{L}^\epsilon_{4}= \epsilon_{\mu
\nu \rho \sigma} \partial^\mu A^\nu \partial^\rho A^ \sigma  =
\frac{1}{2} F_{\mu \nu}\tilde{F}^{\mu \nu}$, which is not a new term,
being included by construction in $\mathcal{L}_2$. At this order, this is
the only possibility with vanishing Hessian constraints.

Cubic terms produce two test Lagrangians, namely

\begin{equation}
\mathcal{L}^\text{test}_{5,1} =  \epsilon_{\mu \nu \rho
\sigma} \partial^\mu A^\nu \partial^\rho A^ \sigma  \left(\partial \cdot
A \right)= \frac{1}{2} F_{\mu \nu} \tilde{F}^{\mu \nu} \left(\partial
\cdot A \right)
\ \ \ \ \hbox{and} \ \ \ \
\mathcal{L}^\text{test}_{5,2}= \epsilon_{\mu \nu \rho \sigma} 
\partial^\mu A ^\nu \partial ^\rho A_\alpha \partial^\alpha A^\sigma = 
\tilde{F}_{\rho \sigma}   \partial ^\rho A_\alpha \partial^\alpha
A^\sigma,
\end{equation}
which, once combined in such a way as to ensure the vanishing of
the Hessian $\mathcal{H}^{0\mu}$, yield
\begin{equation}
\mathcal{L}_{5}^{\epsilon} = F_{\mu \nu} \tilde{F}^{\mu \nu}
\left(\partial \cdot A \right) - 4  \left( \tilde{F}_{\rho \sigma}  
\partial ^\rho A_\alpha \partial^\alpha A^\sigma \right).
\end{equation}

Increasing the powers of the derivatives, as usual by now, also increases
the complexity of the possible new terms. Quartic test terms are found
to be given by
\begin{equation}
\begin{split}
\mathcal{L}^{\epsilon,\text{test}}_{6,1} & = \left( \epsilon_{\mu \nu \rho \sigma} \partial^\mu A^\nu \partial^\rho A^ \sigma \right) ^2 = \frac{1}{4} \left(F_{\mu \nu} \tilde{F}^{\mu \nu}\right)^2, \\
\mathcal{L}^{\epsilon,\text{test}}_{6,2} & = \left( \epsilon_{\mu \nu \rho \sigma} \partial^\mu A^\nu \partial^\rho A^ \sigma \right) \left(\partial \cdot A \right)^2 =   \frac{1}{2} F_{\mu \nu} \tilde{F}^{\mu \nu} \left(\partial \cdot A \right)^2,  \\
\mathcal{L}^{\epsilon,\text{test}}_{6,3} & = \left( \epsilon_{\mu \nu \rho \sigma} \partial^\mu A^\nu \partial^\rho A^ \sigma \right) \left( \partial ^\alpha A^\beta \partial_\alpha A_\beta \right),   \\
\mathcal{L}^{\epsilon,\text{test}}_{6,4} & = \left( \epsilon_{\mu \nu \rho \sigma} \partial^\mu A^\nu \partial^\rho A ^\sigma \right) \left( \partial ^\alpha A^\beta \partial_\beta A_\alpha \right) , \\
\mathcal{L}^{\epsilon,\text{test}}_{6,5} & = \left( \epsilon_{\mu \nu \rho \sigma}  \partial^\mu A ^\nu \partial ^\rho A_\alpha \partial^\alpha A^\sigma \right) \left(\partial \cdot A \right) =  \left( \tilde{F}_{\rho \sigma}   \partial ^\rho A_\alpha \partial^\alpha A^\sigma \right) \left(\partial \cdot A \right),  \\
\mathcal{L}^{\epsilon,\text{test}}_{6,6} & = \left( \epsilon_{\mu \nu \rho \sigma} \partial^\mu A^\nu \partial^\rho A_\alpha \partial^\sigma A_\beta \partial^\alpha A^\beta \right) = \left(  \tilde{F}_{\rho \sigma} \partial^\rho A_\alpha \partial^\sigma A_\beta \partial^\alpha A^\beta \right), \\
\mathcal{L}^{\epsilon,\text{test}}_{6,7}& = \left( \epsilon_{\mu \nu \rho \sigma} \partial^\mu A^\nu \partial_\alpha A^\rho \partial_\beta A^\sigma \partial^\alpha A^\beta \right) =  \left(  \tilde{F}_{\rho \sigma} \partial_\alpha A^\rho \partial_\beta A^\sigma \partial^\alpha A^\beta \right),\\
\mathcal{L}^{\epsilon,\text{test}}_{6,8} & = \left( \epsilon_{\mu \nu \rho \sigma} \partial^\mu A^\nu \partial^\rho A_\alpha \partial_\beta A^\sigma\partial^\alpha A^\beta \right) =  \left(  \tilde{F}_{\rho \sigma} \partial^\rho A_\alpha \partial_\beta A^\sigma \partial^\alpha A^\beta \right), \\
\mathcal{L}^{\epsilon,\text{test}}_{6,9} & =   \left( \epsilon_{\mu \nu \rho \sigma} \partial^\mu A^\nu \partial^\rho A_\beta \partial_\alpha A^\sigma\partial^\alpha A^\beta \right) =  \left(  \tilde{F}_{\rho \sigma} \partial^\rho A_\beta \partial_\alpha A^\sigma \partial^\alpha A^\beta \right),
\end{split}
\end{equation}
leading to one new solution with vanishing Hessian constraint, namely
\begin{equation}
\mathcal{L}_{6}^{\epsilon}=  \tilde{F}_{\rho \sigma} F^\rho{}_\beta
F^\sigma{}_\alpha \partial^\alpha A^\beta \,,
\end{equation}
together with $\mathcal{L}_{F\tilde{F}\times F\tilde{F}}=\left(F_{\mu \nu}
\tilde{F}^{\mu \nu}\right)^2$ and $\mathcal{L}_{F\tilde{F}\times
FF}=\left(F_{\mu \nu} \tilde{F}^{\mu \nu}\right) \left( F_{\rho \sigma}
F^{\rho\sigma} \right)$, both of which are already part of
$\mathcal{L}_2$. As before, all these extra Lagrangians can be multiplied
by arbitrary functions $f^\epsilon (X)$ without modifying our conclusions.

\subsection{Final Lagrangian in flat spacetime}

Up to this point, we have considered explicit terms involving products
of up to five derivatives of the vector field. Higher-order terms can be
derived by continuing along the same lines of calculation; this would
imply an important number of long terms. We have not found a general
rule allowing to derive a generic action at any given order, so we are
merely led to conjecture, given the above calculations, that there is no
reason the higher-order terms to be vanishing (assuming they do not
lead to trivially vanishing equations of motion \cite{Deffayet:2013tca}).

Considering the result previously derived, we conjecture that terms like
$\mathcal{L}_{n}^{\text{Perm}}$ will continue to show up at higher
order, and in fact, we speculate that the higher order the more numerous
terms one will find. A generic $\mathcal{L}_{n}^{\text{Perm}}$ will
yield vanishing $\mathcal{H}^{00}$ and $\mathcal{H}^{0i}$ and vanish
when going to the scalar limit $A_\mu\to\partial_\mu\pi$ of the theory.
Including the Levi-Civita terms, we propose that the most general
Lagrangian leading to second-order equations of motion for three vector
propagating degrees of freedom contains an infinite number of terms,
whose general formulation takes the form
\begin{equation}
\label{LagFinalSum}
\mathcal{L}_{\text{gen}}\left(A_\mu\right)=-\frac{1}{4}F_{\mu\nu}F^{\mu\nu}+\sum_{n\geq
  2} \mathcal{L}_n + \sum_{n\geq 5}\mathcal{L}^\epsilon_n,
\end{equation}
with
\begin{equation}
\label{LagFinal}
\begin{split}
\mathcal{L}_2 = & f_2(A_\mu,F_{\mu\nu},\tilde{F}_{\mu\nu}),\\
\mathcal{L}_3 = & f_3^{\text{Gal}}(X) \mathcal{L}_3^{\text{Gal}},\\
\mathcal{L}_4 = & f_4^{\text{Gal}}(X) \mathcal{L}_4^{\text{Gal}},\\
\mathcal{L}_5 = & f_5^{\text{Gal}}(X) \mathcal{L}_5^{\text{Gal}}+f_5^{\text{Perm}}(X)
\mathcal{L}_5^{\text{Perm}},\\
\mathcal{L}_6 = &f_6^{\text{Perm}}(X) \mathcal{L}_6^{\text{Perm}},\\
\mathcal{L}_7 = & f_7^{\text{Perm,1}}(X) \mathcal{L}_7^{\text{Perm,1}}
+ f_7^{\text{Perm,2}}(X) \mathcal{L}_7^{\text{Perm,2}},\\
\mathcal{L}_{n\geq 8} = & \sum_{i} f_n^{\text{Perm},i}(X) \mathcal{L}_n^{\text{Perm},i},\\
\mathcal{L}^\epsilon_n = & \sum_{i} g_n^{\epsilon,i}(X) \mathcal{L}_n^{\epsilon,i},
\end{split}
\end{equation}
where the terms in $\mathcal{L}_3$ to $\mathcal{L}_5$ are given in
Sec. \ref{L3->5}, those in $\mathcal{L}_6$ and $\mathcal{L}_7$
can be found respectively in Sec. \ref{L6} and \ref{L7}, and those in $\mathcal{L}^\epsilon_n$ are shown in Sec. \ref{sec:eps}. Once again,
$f_2$ is an arbitrary scalar function of
all the possible contractions among $A_\mu$, $F_{\mu\nu}$ and
$\tilde{F}_{\mu\nu}$, and all the other $f_n$ and $g_n^{\epsilon}$
are arbitrary functions of $X$.

\section{Curved space-time}
\label{Covariantization}

Coupling the vector with the metric and rendering the latter dynamical
implies many new possible terms satisfying our conditions. Those are
explored below. Some of the terms have already been discussed in
previous works \cite{Horndeski:1976gi,Heisenberg:2014rta,Hull:2015uwa},
and we introduce new ones below.

\subsection{Covariantization}

In order to take into account the metric $g_{\mu\nu}$ itself as
dynamical in the most general way, one needs to impose its equation of
motion to be also of order two at most. First, we transform partial
derivatives into covariant derivatives. This satisfies the constraints
in the vector sector since the additional terms, coming from the
connection, are only first order in the metric derivatives and thus
ensures that no derivative higher than second order will appear in the
equations of motion of the vector field or the metric.

On the other hand, when going to the scalar sector, one has to pay
special attention to the higher derivative terms which can appear due to
the commutations of derivatives of the scalar field. This can be
problematic only when one considers terms reducing to the scalar
Galileon Lagrangians because the property $\nabla_\mu\nabla_\nu
\pi=\nabla_\nu\nabla_\mu \pi$ remains valid in curved spacetime. As for
the extra terms, called $\mathcal{L}_n^{\text{Perm}}$ above, as well as
the U(1)-invariants, they keep vanishing in the scalar formulation of
the theory. The terms reducing to the scalar Galileon Lagrangians have
already been studied, and their curved space-time extension can be found
in Refs.~\cite{Deffayet:2009wt,Deffayet:2011gz,Heisenberg:2014rta}. In
conclusion, the only terms to be modified in Eq. (\ref{LagFinalSum}) are
\begin{equation}
\label{LagFinal}
\begin{split}
\mathcal{L}_4 & = f_4^{\text{Gal}}(X)R -2 f_{4,X}^{\text{Gal}}(X)
\mathcal{L}_4^{\text{Gal}},\\
\mathcal{L}_5 & =f_5^{\text{Gal}}(X)
G_{\mu\nu} \nabla^\mu A^\nu+3 f_{5,X}^{\text{Gal}}(X)
\mathcal{L}_5^{\text{Gal}}+f_5^{\text{Perm}}(X)
\mathcal{L}_5^{\text{Perm}},\\
\end{split}
\end{equation}
where the notation $f_{,X}$ stands for a derivative with respect to $X$,
i.e. $f_{,X}\equiv \dd f/\dd X$.

\subsection{Additional curvature terms}

This last part is dedicated to all the additional terms which can appear from
the coupling contractions of curvature terms with terms implying the vector
field. A similar study was already proposed for the Abelian case, and it was
shown that the only possibility was to contract the field tensor with
divergence-free objects built from curvature
\cite{deRham:2011by,Jimenez:2013qsa}, i.e. the Einstein tensor
$G_{\mu\nu}$ and the following fourth-rank divergence-free tensor
\begin{equation}
L_{\mu\nu\rho\sigma}=2R_{\mu\nu\rho\sigma} +
2(R_{\mu\sigma}g_{\rho\nu} + R_{\rho\nu}g_{\mu\sigma} -
R_{\mu\rho}g_{\nu\sigma} - R_{\nu\sigma}g_{\mu\rho}) +
R(g_{\mu\rho}g_{\nu\sigma} - g_{\mu\sigma}g_{\rho\nu}).
\end{equation}
This tensor has the same symmetry properties as the Riemann tensor
$R_{\mu\nu\rho\sigma}$, i.e. it is antisymmetric in $(\mu,\nu)$ and
$(\rho,\sigma)$, and symmetric in the exchange of ($\mu\nu)$ and
($\rho\sigma$). The use of divergence-free tensors permits us to avoid higher-order derivatives of the metric in the equations of motion: contracting
such a tensor with another dynamical object of the required derivative
order will naturally lead, in the equations of motion, to the divergence
of this tensor, and hence will vanish in the chosen case.

In order to derive the relevant generalizing terms for the Proca theory,
we proceed in the same way as in the previous section. We first consider
contractions of both $G_{\mu\nu}$ and $L_{\mu\nu\rho\sigma}$ with
$A_\mu$ only, and then these specific contractions with $A_\mu$ and its
first derivative that vanish in the scalar part. We then apply the
Hessian condition given in Sec. \ref{procedure} to obtain the required
terms.

For the contractions with the vector field only, the only term we can
have is
\begin{equation}
\mathcal{L}^\text{Curv}_1=G_{\mu\nu}A^\mu A^\nu,
\end{equation}
with all possible contractions with $L_{\mu\nu\rho\sigma}$ being vanishing due to
its antisymmetry properties. Note that in this case, we cannot multiply
this Lagrangian by a scalar function of the vector field $X$: in the
scalar sector, this would lead to first-order derivatives of the scalar
component $\pi$ which would subsequently yield terms involving
$\nabla_{\alpha} G_{\mu\nu}$ in the equations of motion, terms which
are third order in the metric and thus excluded.

Contractions with the field tensor only have already been studied in
Ref.~\cite{Jimenez:2013qsa}, where it was shown that the only available
term satisfying our requirements can be written as
\begin{equation}
\mathcal{L}^\text{Curv}_2=L_{\mu\nu\rho\sigma} F^{\mu\nu} F^{\rho\sigma},
\label{L2curv}
\end{equation}
the other possible contraction over different indices,
$L_{\mu\nu\rho\sigma}F^{\mu\rho}F^{\nu\sigma}$, being equivalent to
Eq.~\eqref{L2curv} because of the symmetries of $L_{\mu\nu\rho\sigma}$.
Similarly, one could consider the term built out of the dual of the
field tensor;  this actually also reduces to $\mathcal{L}^\text{Curv}_2$
\cite{Jimenez:2013qsa}. Moreover, this last term can be multiplied by
any scalar function of the vector field only without losing its
properties, since it vanishes in the scalar sector.

Last on the list are those non-U(1) invariant terms with first-order
derivatives of the vector field; these must vanish in the scalar sector
in order to comply with our demands. In this case, in order to have at
most a second-order equation of motion of the metric from the equation of
motion of the vector field, it is sufficient to contract only the
indices of the derivatives with the divergence-free curvature tensors.
At this point, we see that we need to introduce a term in
$\nabla^{\mu}A^{\nu}$, if we want to commute derivatives in order to have
a vanishing term in the scalar sector, and that this term will thus
involve the field strength tensor. As a result, we can consider terms
with either $L_{\mu\nu\alpha\beta} F^{\mu\nu}$ or
$L_{\mu\alpha\rho\beta} F^{\mu\rho}$ contracted with $(\nabla^\alpha
A^\beta)$, $(\nabla^\beta A^\alpha)$, $(\nabla^\alpha A^\gamma
\nabla^\beta A_\gamma)$ or $(A^\alpha A^\beta)$. A close examination of
these possibilities reveals that such terms are either vanishing, or
else that they reduce to $\mathcal{L}^\text{Curv}_2$.

We summarize the set of all possible additional terms containing the
curvature in the following Lagrangian
\begin{equation}
\mathcal{L}^\text{Curv}= f^\text{Curv}_1 G_{\mu\nu}A^\mu A^\nu +
f^\text{Curv}_2(X)L_{\mu\nu\rho\sigma} F^{\mu\nu} F^{\rho\sigma},
\end{equation}
where $f^\text{Curv}_1$ is a constant, and $f^\text{Curv}_2$ is an
arbitrary function of $X$ only. We did not consider higher-order
products of curvature terms since our aim is to focus on the the vector
part of the theory.

\subsection{Final Lagrangian in curved spacetime}

We can finally write the complete expression of the generalized
Abelian Proca theory in curved spacetime, which reads
\begin{equation}
\label{LagFinalSumCST}
\mathcal{L}_{\text{gen}}=-\frac{1}{4}F_{\mu\nu}F^{\mu\nu}
+\mathcal{L}^{\text{Curv}}+\sum_{n\geq 2} \mathcal{L}_n
+ \sum_{n\geq 5}\mathcal{L}^\epsilon_n,
\end{equation}
where
\begin{equation}
\label{LagFinalCST}
\begin{split}
\mathcal{L}^\text{Curv} & = f^\text{Curv}_1 G_{\mu\nu}A^\mu A^\nu +
f^\text{Curv}_2(X)L_{\mu\nu\rho\sigma} F^{\mu\nu} F^{\rho\sigma},\\
\mathcal{L}_2 & =f_2(A_\mu, F_{\mu\nu},\tilde{F}_{\mu\nu}),\\
\mathcal{L}_3 & = f_3^{\text{Gal}}(X) \mathcal{L}_3^{\text{Gal}},\\
\mathcal{L}_4 & = f_4^{\text{Gal}}(X)R -2 f_{4,X}^{\text{Gal}}(X)
\mathcal{L}_4^{\text{Gal}},\\
\mathcal{L}_5 & =f_5^{\text{Gal}}(X) G_{\mu\nu} \nabla^\mu A^\nu
+3 f_{5,X}^{\text{Gal}}(X) \mathcal{L}_5^{\text{Gal}}+f_5^{\text{Perm}}(X)
\mathcal{L}_5^{\text{Perm}},\\
\mathcal{L}_6 & =f_6^{\text{Perm}}(X) \mathcal{L}_6^{\text{Perm}},\\
\mathcal{L}_7 & =f_7^{\text{Perm,1}}(X) \mathcal{L}_7^{\text{Perm,1}}+
f_7^{\text{Perm,2}}(X) \mathcal{L}_7^{\text{Perm,2}},\\
\mathcal{L}_{n\geq 8} & =\sum_{i} f_n^{\text{Perm},i}(X) \mathcal{L}_n^{\text{Perm},i},\\
\mathcal{L}^\epsilon_n & = \sum_{i} g_n^{\epsilon,i}(X) \mathcal{L}_n^{\epsilon,i},
\end{split}
\end{equation}
all $f$ and $g$ being arbitrary functions of $X$, except $f^\text{Curv}_1$
which is a constant, and $f_2$ which is an
arbitrary function of $A_\mu$, $F_{\mu\nu}$ and $\tilde{F}_{\mu\nu}$. The complete
expression of the Lagrangians which appears in Eq. (\ref{LagFinalCST})
are given in Sec. \ref{L3->5} to \ref{sec:eps} where
partial derivatives must be replaced by covariant ones.

As in the previous, flat-space case, we conjecture the series to contain an
infinite number of terms.

\section{Conclusion and discussion}

In this paper, we investigated the terms which can be included in the
generalized Proca theory. After obtaining those which contain products
of up to five first derivatives, we proposed a conjecture concerning the
following terms. We discuss a crucial difference between the generalized
scalar and vector theories: in contrast with the Galileon theory, the
generalization of the Proca theory seems to depend on an infinite number
of arbitrary functions, either in front of
$\mathcal{L}^{\text{Perm},i}_n$ or in front of all the possible
contractions among field tensors.

On the other hand, one could ask if this theory is the most general one
can write, in flat as in curved  spacetime\footnote{The most general
multiple-scalar field theory in flat spacetime was given in Ref.
\cite{Sivanesan:2013tba} and covariantized in Ref.
\cite{Padilla:2012dx}.  In the latter paper, it was suggested that the
ensuing theory would be the multi-scalar generalization of Horndeski's
theory.  However, as shown in Ref. \cite{Kobayashi:2013ina}, the
covariantization procedure does not guarantee that the resulting action
is the most general;  indeed, as presented in Ref.
\cite{Ohashi:2015fma}, a better procedure is that devised originally
by Horndeski in Ref. \cite{Horndeski:1974wa}.}. A first extension would,
for instance, consist in including terms containing higher-order
derivatives leading still to second-order equations of motion; this can
be found for instance in Ref.~\cite{Deffayet:2010zh}.  A second
extension would be to investigate the terms additional to those
appearing in the covariantization of the action, through the equation of
motion rather than the Lagrangian, as it is done in Ref.
\cite{Ohashi:2015fma} following the original Horndeski's procedure
\cite{Horndeski:1974wa}; this could be interesting in order to have a
full understanding of the model.

In the future, this work could be completed along many directions. For
instance, a Hamiltonian analysis of the relevant degrees of freedom
would allow us to determine the relevant symmetries. Moreover, one
could study how a general theory such as that defined here can be obtained
from an effective U(1) invariant initial model. Pioneering work on this
direction, along the line of spontaneous symmetry breaking to obtain
extensions of Proca theory, has been done and can be found in, e.g.,
Refs.~\cite{Tasinato:2014mia,Hull:2014bga}. An extension to
the non-Abelian situation is currently undergoing, with a much richer
phenomenology. Finally, one could consider the effects of such a theory in a
cosmological context, be it by means of implementing an inflation model 
\cite{Jimenez:2013qsa,Barrow:2012ay} or by suggesting new solutions to
the cosmological constant problem \cite{Tasinato:2014mia}.

\section*{Acknowledgments} Throughout this paper, we have used xTensor,
xCoba and TexAct, all parts of the Mathematica package xAct available in
Ref.~\cite{xAct}. We wish to thank L.~Bernard, C.~Deffayet,
G.~Esposito-Far\`ese, L.~Heisenberg and  G.~Faye for illuminating
discussions.  This work was supported by COLCIENCIAS - ECOS NORD grant
number RC 0899-2012 with the help of ICETEX, and by COLCIENCIAS grant
numbers 110656933958 RC 0384-2013 and 123365843539 RC FP44842-081-2014.

\appendix
\section{Hessian condition for the quartic terms: $\mathcal{L}_6$}
\label{appL6}

The different test Lagrangians that can be written down at the
fourth-order derivative level are
\begin{equation}
\begin{split}
\mathcal{L}^\text{test}_{6,1} & = \left(\partial\cdot A\right)^4,\\
\mathcal{L}^\text{test}_{6,2} & = \left(\partial\cdot A\right)^2
   \left(\partial_ {\sigma}A_ {\rho} \partial^ {\sigma}A^ {\rho}\right),\\
\mathcal{L}^\text{test}_{6,3} & = \left(\partial\cdot A\right)^2
   \left(\partial_ {\rho}A_ {\sigma} \partial^ {\sigma}A^{\rho}\right),\\
\mathcal{L}^\text{test}_{6,4} & = \left(\partial\cdot A\right) \left(\partial_\nu
    A_ {\sigma} \partial^ {\rho}A^{\nu} \partial^{\sigma} A_ {\rho}\right),\\
\mathcal{L}^\text{test}_{6,5} & = \left(\partial\cdot A\right) \left(\partial^ {\rho}A^ {\nu}
   \partial_ {\sigma}A_ {\rho} \partial^ {\sigma}A_\nu\right),\\
\mathcal{L}^\text{test}_{6,6} & = \left(\partial_\mu A_ {\sigma} \partial^{\nu} A^ {\mu}
   \partial^\rho A_ {\nu} \partial^ {\sigma}A_\rho\right),\\
\mathcal{L}^\text{test}_{6,7} & = \left(\partial^\nu A^ {\mu} \partial_{\rho} A_ {\sigma}
   \partial^\rho A_ {\mu} \partial^ {\sigma}A_\nu \right),\\
\mathcal{L}^\text{test}_{6,8} & = \left(\partial_\nu A^ {\sigma} \partial^{\nu} A^{\mu}
   \partial_\rho A_ {\sigma} \partial^ {\rho}A_\mu\right),\\
\mathcal{L}^\text{test}_{6,9} & = \left(\partial^\nu A^ {\mu} \partial^ {\rho}A_ {\mu}
   \partial_\sigma A_ {\rho} \partial^ {\sigma}A_\nu \right),\\
\mathcal{L}^\text{test}_{6,10} & = \left(\partial_\nu A_ {\mu} \partial^ {\nu}A^ {\mu}\right)^2,\\ \mathcal{L}^\text{test}_{6,11} & = \left(\partial_{\mu}A_\nu \partial^ {\nu}A^ {\mu}\right)
   \left(\partial_ {\sigma}A_\rho \partial^ {\sigma}A^\rho \right),\\
\mathcal{L}^\text{test}_{6,12} & = \left(\partial_ {\mu}A_ {\nu} \partial^ {\nu}A^\mu \right)
   \left( \partial_ {\rho}A_{\sigma} \partial^ {\sigma}A^{\rho}\right),
\end{split}
\end{equation}
so that setting
\begin{equation}
\mathcal{L}^\text{test}_{6} = \sum_{k=1}^{12} x_k \, \mathcal{L}^\text{test}_{6,k},
\end{equation}
we obtain for the $(00)$ component of the Hessian
\begin{equation}
\begin{split}
\mathcal{H}_{6}^{00} = & 2 \left(6 x_{1} + x_{2} + x_{3}\right)
\left(\partial \cdot A\right)^2 + 2 \left(2 x_{10} + x_{11} +
x_{2}\right) \left(\partial_ {\mu}A_ {\nu} \partial^ {\mu}A^
{\nu}\right) \\
& + 2 \left(x_{3}+x_{11} + 2 x_{12}\right) \left(\partial^
{\mu}A^ {\nu} \partial_ {\nu}A_ {\mu}\right) - 2 \bigl[4 x_{2} + 4
  x_{3} + 3 \left(x_4 + x_5\right) \bigr] \left(\partial \cdot
A\right) \left(\partial^{0}A^{0}\right) \\ & + 4 \left[ x_6 + x_7 + x_8 + x_9
+2 \left( x_{10} + x_{11} + x_{12}\right) \right]
\left(\partial^{0}A^{0}\right)^2 - 2 \left( x_5 + x_7 + 2
x_8 + x_9\right) \left(\partial_ {\mu}A^{0}\right) \left(\partial^
{\mu}A^{0}\right) \\
& - \left(\partial^{0}A_ {\mu}\right) \left\{2 \left(
x_5 + x_7 + 2 x_8 + x_9\right) \left(\partial^{0}A^
{\mu}\right) + 2 \bigl[3 x_4 + x_5 + 2 \left(2 x_6 + x_7 +
  x_9\right) \bigr] \left(\partial^ {\mu}A^{0}\right)\right\},
\end{split}
\end{equation}
and the $(0i)$ component
\begin{equation}
\begin{split}
\mathcal{H}_{6}^{0i} = & 2 \left(x_7 + 2 x_8 + x_9 + 4 x_{10} + 2 x_{11} \right)
\left(\partial^{0}A^{0}\right)
\left(\partial^{0}A^{i}\right) + 2 \left(2 x_6 + x_7 + x_9 +2 x_{11} + 4 x_{12} \right)
\left(\partial^{0}A^{0}\right)
\left(\partial^{i}A^{0}\right) \\
& - \left(\partial \cdot A\right)
\left[2 \left(2 x_{2} + x_5\right) \left(\partial^{0}A^{i}\right) +
  \left(4 x_{3} + 3 x_4 + x_5\right)
  \left(\partial^{i}A^{0}\right) \right] - \left(3 x_4 + 4 x_6 +
x_7\right) \left(\partial^{i}A_ {\mu}\right)\left( \partial^
{\mu}A^{0}\right) \\
& - \left( x_5 + x_7 + 4 x_8\right)
\left(\partial_ {\mu}A^{i}\right)\left( \partial^ {\mu}A^{0}\right) -
\left(\partial^{0}A_ {\mu}\right) \left(x_5 + x_7 + 2 x_9\right)
\left( \partial^{i}A^ {\mu} + \partial^ {\mu}A^{i}\right).
\end{split}
\end{equation}
Canceling these two functions of $x_k$ provides the solutions 
exhibited in Sec.~\ref{L6}.

\section{Hessian condition for the fifth order: $\mathcal{L}_7$}
\label{appL7}

In the last case we considered, for which the highest power in derivatives
is five, we can write the various test Lagrangians as
\begin{equation}
\begin{split}
\mathcal{L}^\text{test}_{7,1} & = \left(\partial\cdot A\right)^5,\\ 
\mathcal{L}^\text{test}_{7,2} & = \left(\partial\cdot A\right) ^3
\left(\partial_ {\gamma}A_ {\sigma} \partial^ {\gamma}A^{\sigma}\right),\\
\mathcal{L}^\text{test}_{7,3} & = \left(\partial\cdot A\right) ^3
   \left(\partial_ {\sigma}A_ {\gamma} \partial^{\gamma} A^{\sigma}\right),\\
\mathcal{L}^\text{test}_{7,4} & = \left(\partial\cdot A\right)^2\left(\partial_ {\rho}A_ {\gamma}
   \partial^ {\sigma}A^ {\rho} \partial^ {\gamma}A_{\sigma}\right),\\
\mathcal{L}^\text{test}_{7,5} & = \left(\partial\cdot A\right) ^2 \left(\partial^{\sigma}
   A^ {\rho} \partial_ {\gamma}A_ {\sigma} \partial^ {\gamma}A_{\rho}\right),\\
\mathcal{L}^\text{test}_{7,6} & = \left(\partial\cdot A\right) \left(\partial_{\nu} A_ {\gamma}
   \partial^ {\rho}A^{\nu} \partial^ {\sigma}A_ {\rho} \partial^ {\gamma}A_{\sigma}\right),\\
\mathcal{L}^\text{test}_{7,7} & = \left(\partial\cdot A\right) \left(\partial^ {\rho}A^ {\nu}
   \partial_ {\sigma}A_ {\gamma}\partial^ {\sigma}A_ {\nu}\partial^ {\gamma}A_{\rho}\right),\\
\mathcal{L}^\text{test}_{7,8} & = \left(\partial\cdot A\right)\left(\partial_{\rho}A^ {\gamma}
   \partial^ {\rho}A^{\nu}\partial_{\sigma}A_{\gamma}\partial^{\sigma}A_{\nu}\right),\\
\mathcal{L}^\text{test}_{7,9} & = \left(\partial\cdot A\right) \left(\partial^ {\rho}A^{\nu}
   \partial^{\sigma}A_{\nu}\partial_{\gamma}A_{\sigma}\partial^{\gamma}A _{\rho}\right),\\
\mathcal{L}^\text{test}_{7,10} & = \left(\partial\cdot A\right)\left(\partial_{\rho}A_{\nu}
   \partial^{\rho}A^{\nu}\right)^2,\\
\mathcal{L}^\text{test}_{7,11} & = \left(\partial\cdot A\right) \left(\partial_
{\nu}A_ {\rho} \partial^ {\rho}A^ {\nu}\right) \left(\partial_
{\gamma}A_ {\sigma} \partial^ {\gamma}A^ {\sigma}\right),\\
\mathcal{L}^\text{test}_{7,12} & = \left(\partial\cdot A\right) \left(\partial_
{\nu}A_ {\rho} \partial^ {\rho}A^ {\nu}\right)\left( \partial_
{\sigma}A_ {\gamma} \partial^ {\gamma}A^ {\sigma}\right),\\
\mathcal{L}^\text{test}_{7,13} & = \left(\partial_ {\mu}A_ {\gamma} \partial^ {\nu}A^
{\mu} \partial^ {\rho}A_ {\nu} \partial^ {\sigma}A_ {\rho} \partial^
{\gamma}A_ {\sigma}\right),\\
\mathcal{L}^\text{test}_{7,14} & = \left(\partial^
{\nu}A^ {\mu} \partial_ {\rho}A_ {\gamma} \partial^ {\rho}A_ {\mu}
\partial^ {\sigma}A_ {\nu} \partial^ {\gamma}A_ {\sigma}\right),\\
\mathcal{L}^\text{test}_{7,15} & = \left(\partial^ {\nu}A^ {\mu} \partial^ {\rho}A_
{\mu} \partial^ {\sigma}A_ {\nu} \partial_ {\gamma}A_ {\sigma} \partial^
{\gamma}A_ {\rho}\right),\\
\mathcal{L}^\text{test}_{7,16} & = \left(\partial^ {\nu}A^
{\mu} \partial_ {\rho}A^ {\gamma} \partial^ {\rho}A_ {\mu} \partial_
{\sigma}A_ {\gamma} \partial^ {\sigma}A_ {\nu}\right),\\
\mathcal{L}^\text{test}_{7,17} & = \left(\partial_ {\nu}A_ {\mu} \partial^ {\nu}A^
{\mu}\right)\left( \partial_ {\rho}A_ {\gamma} \partial^ {\sigma}A^
{\rho} \partial^ {\gamma}A_ {\sigma}\right),\\
\mathcal{L}^\text{test}_{7,18} & =
\left(\partial_ {\nu}A_ {\mu} \partial^ {\nu}A^ {\mu} \right)
\left(\partial^ {\sigma}A^ {\rho} \partial_ {\gamma}A_ {\sigma}
\partial^ {\gamma}A_ {\rho}\right),\\
\mathcal{L}^\text{test}_{7,19} & =
\left(\partial_ {\mu}A_ {\nu} \partial^ {\nu}A^ {\mu}
\right)\left(\partial_ {\rho}A_ {\gamma} \partial^ {\sigma}A^ {\rho}
\partial^ {\gamma}A_ {\sigma}\right),\\
\mathcal{L}^\text{test}_{7,20} & =
\left(\partial_ {\mu}A_ {\nu} \partial^ {\nu}A^ {\mu} \right)
\left(\partial^ {\sigma}A^ {\rho} \partial_ {\gamma}A_ {\sigma}
\partial^ {\gamma}A_ {\rho}\right).
\end{split}
\end{equation}
We again set
\begin{equation}
\mathcal{L}^\text{test}_{6} = \sum_{k=1}^{20} x_k \, \mathcal{L}^\text{test}_{7,k},
\end{equation}
leading to the following Hessian matrix elements: 
\begin{equation}
\begin{split}
\mathcal{H}_{7}^{00} = & 2 \left(10 x_{1} + x_{2} + x_{3}\right)
\left(\partial \cdot A\right)^3 + 2 \left(x_{4} + x_{17} + x_{19}\right)
\left(\partial^ {\mu}A^ {\nu} \partial_ {\nu}A_ {\rho} \partial^
{\rho}A_ {\mu}\right) \\ 
& + 2 \left(x_{5} + x_{18} + x_{20} \right) \left(\partial^
{\mu}A^ {\nu} \partial_ {\rho}A_ {\mu} \partial^ {\rho}A_ {\nu}\right)
 - 6 \left(2 x_{2} + 2 x_{3} + x_{4} + x_{5}\right) \left(\partial \cdot
A\right)^2 \left(\partial^{0}A^{0}\right) \\
& - 2 \bigl[4 x_{10} + 2 x_{11} + 3 \left(x_{17} + x_{18}\right) \bigr] 
\left(\partial_ {\mu}A_ {\nu} \partial^ {\mu}A^ {\nu}\right) \left(\partial^{0}A^{0}\right) \\
& - 2 \bigl[2 x_{11} + 4 x_{12} + 3 \left(x_{20} + x_{19}\right) \bigr] \left(\partial^
{\mu}A^ {\nu} \partial_ {\nu}A_ {\mu}\right) \left(\partial^{0}A^{0}
\right) \\
& + \left(\partial \cdot A\right) \bigl\{2 \left(3 x_{2}+ 2 x_{10} + x_{11} 
\right) \left(\partial_ {\mu}A_ {\nu} \partial^ {\mu}A^
{\nu}\right) + 2 \left(3 x_{3} + x_{11} + 2 x_{12} \right)
\left(\partial^ {\mu}A^ {\nu} \partial_ {\nu}A_ {\mu}\right) \\
& + 4 \left[ x_{6} + x_{7} + x_{8} + x_{9} + 2 \left(
x_{10} + x_{11} + x_{12} \right) \right]
\left(\partial^{0}A^{0}\right)^2 \\
& - 2 \left(2 x_{5} + x_{7} + 2 x_{8}
+ x_{9}\right) \left(\partial_ {\mu}A^{0}\right)\left(\partial^
{\mu}A^{0}\right) - \left(\partial^{0}A_ {\mu}\right) \bigl[2 \left(2
x_{5} + x_{7} + 2 x_{8} + x_{9}\right) \left(\partial^{0}A^ {\mu}
\right)\\
& - 4 \left(3 x_{4} + x_{5} + 2 x_{6} + x_{7} + x_{9}\right)
\left(\partial^ {\mu}A^{0}\right) \bigr]\bigr\} \\
& +
\left(\partial^{0}A^{0} \right)\bigl\{2 \left[ x_{14} + x_{15} + 2 \left(
x_{16} + x_{18} + x_{20} \right)\right] \left[ \left( \partial_ {\mu}A^{0}\right)\left( \partial^
{\mu}A^{0}\right)  + \left(\partial^{0}A_ {\mu}\right) \left(\partial^{0}A^ {\mu}
\right) \right] \\
& + 2 \left(5 x_{13} + 3 x_{14} + 3 x_{15} + x_{16} + 6 x_{17} + 2 x_{18} + 6 x_{19} + 2 x_{20} \right)
\left( \partial^ {\mu}A^{0}\right) \left( \partial^0 A_\mu\right) \bigr]\bigr\} \\
& - \left(\partial^ {\mu}A^ {\nu}\right) \left(\partial^{0}A_ {\mu}
\right)\bigl\{2 \bigl[  x_{7} + x_{14} + x_{16} + 2 \left(x_{9} +
x_{15}\right) \bigr]\left( \partial^{0}A_ {\nu} \right) \\
& - 2 \left(4 x_{6} + x_{7} + 5 x_{13} + 2 x_{14} + x_{15}\right) \left(\partial_
{\nu}A^{0}\right)\bigr\} \\
& -\left(\partial^ {\mu}A^ {\nu}\right) \left(
\partial_ {\mu}A^{0} \right)\bigl\{2 \left(x_{7} + 4 x_{8} + x_{14} + 3 x_{16} \right)
\left(\partial^{0}A_ {\nu} \right) + 2 \bigl[x_{7} + x_{14} + x_{16} + 2 \left(x_{9} + x_{15}
\right) \bigr] \left(\partial_{\nu}A^{0}\right)\bigr\},
\end{split}
\end{equation}
and
\begin{equation}
\begin{split}
\mathcal{H}_{7}^{0i} = &\left(x_{14} + x_{16} + 2 x_{20} \right)
\left(\partial^{0}A_ {\mu}\right)\left( \partial^{0}A^ {\mu}
\right)\left(\partial^{i}A^{0} \right)\\
& - \left(\partial \cdot A\right)^2 \bigl[2 \left(3 x_{2} + x_{5}\right)
  \left(\partial^{0}A^{i} \right) + \left(6 x_{3} + 3 x_{4} + x_{5}\right)
  \left(\partial^{i}A^{0}\right)\bigr] \\ 
& + \left(\partial_ {\mu}A_ {\nu} \partial^ {\mu}A^ {\nu}\right)
   \bigl[2 \left(2 x_{10} + x_{18}\right) \left(\partial^{0}A^{i} \right)
   + \left(2 x_{11} + 3 x_{17} + x_{18}\right) \left(\partial^{i}A^{0}\right) \bigr] \\
&  + \left(\partial^ {\mu}A^ {\nu} \partial_ {\nu}A_ {\mu}\right) \bigl[2
  \left(x_{11} + x_{20}\right) \left(\partial^{0}A^{i} \right)  +
  \left(4 x_{12} + 3 x_{19} + x_{20}\right) \left(\partial^{i}A^{0}\right)
  \bigr] \\
& + \left( 5 x_{13} + x_{14} + 2 x_{15}+ 6 x_{19} + 2 x_{20} \right)
\left(\partial^{0}A_ {\mu}\right)\left( \partial^{i}A^{0}
\right)\left(\partial^ {\mu}A^{0} \right) \\ 
&  + \left(x_{14} + x_{15} + 2 x_{20} \right) \left(\partial^{i}A^{0} \right)\left(\partial_
{\mu}A^{0}\right)\left( \partial^ {\mu}A^{0} \right)+
\left(\partial^{0}A^{i} \right)\bigl[\left(2 x_{16} + 2 x_{18}\right)
  \left(\partial^{0}A_ {\mu} \right)\left(\partial^{0}A^ {\mu} \right)  \\ 
&  + \left(2 x_{14} + x_{15} + x_{16} + 6 x_{17} + 2 x_{18} \right)\left(
  \partial^{0}A_ {\mu} \right)\left(\partial^ {\mu}A^{0} \right)+
  \left(x_{15} + x_{16} + 2 x_{18} \right)\left( \partial_ {\mu}A^{0}
  \right)\left(\partial^ {\mu}A^{0} \right)\bigr] \\ 
&  +  \left(\partial
\cdot A\right) \bigl\{\left( 2 x_{7} + 4 x_{8} + 2 x_{9} + 8 x_{10} + 4 x_{11}\right)
\left(\partial^{0}A^{0} \right)\left(\partial^{0}A^{i}\right) \\
& + \left(4 x_{6} + 2 x_{7} + 2 x_{9} + 4 x_{11} + 8 x_{12} \right)
\left(\partial^{0}A^{0}\right)\left( \partial^{i}A^{0} \right) \\ 
&  - \left(6 x_{4} + 4 x_{6} + x_{7}\right)
\left(\partial^{i}A^ {\mu} \right)\left(\partial_ {\mu}A^{0}\right) -
\left(2 x_{5} + x_{7} + 4 x_{8}\right) \left(\partial_ {\mu}A^{0}
\partial^ {\mu}A^{i} \right)\\
&- \left(\partial^{0}A_ {\mu} \right)
   \bigl[\left(2 x_{5} + x_{7} + 2 x_{9}\right)
  \left(\partial^{i}A^ {\mu} \right)- \left(2 x_{5} + x_{7} + 2 x_{9}\right)
  \left(\partial^ {\mu}A^{i} \right)\bigr]\bigr\} \\
& + \left(\partial^{0}A^{0} \right)\Bigl[\left(5 x_{13} + 2 x_{14} + x_{15}  + 6 x_{17} + 6 x_{19}\right) \left(\partial^{i}A^{\mu}\right)\left( \partial_ {\mu}A^{0} \right)\\
& + \left(x_{14} + 3 x_{16} + 2 x_{18} + 2 x_{20}
\right) \left(\partial_ {\mu}A^{0}\right)
\left(\partial^ {\mu}A^{i} \right)+ \left(\partial^{0}A_ {\mu}
\right)\bigl\{\left(x_{14} + 2 x_{15} + x_{16} + 2 x_{18} + 2 x_{20} \right)
\left(\partial^{i}A^ {\mu}\right) \\
& + \bigl[x_{14} + x_{15} + x_{16} + 2 \left(x_{18} + 2 x_{20} \right) \bigr]
\left(\partial^{\mu}A^{i}\right)\bigr\}\Bigr] - \left( \partial^ {\mu}A^ {\nu}
\right)\bigl\{\left( x_{7} + x_{14} + x_{16} \right)
\left(\partial^{i}A_ {\nu} \right)\left(\partial_ {\mu}A^{0} \right) \\  
& + \left(\partial^{0}A_ {\nu} \right)\bigl[\left( x_{7} + x_{14} + x_{15}\right)
\left(\partial^{i}A_ {\mu} \right)+ 2 \left(2 x_{8} + x_{16}\right)
\left( \partial_ {\mu}A^{i}\right) \bigr] + \left(4 x_{6} + 5 x_{13} + x_{14}\right)
\left(\partial^{i}A_ {\mu}\right)\left(\partial_ {\nu}A^{0} \right) \\  
& + \left( x_{7} + x_{14} + x_{16} \right)
\left(\partial_ {\mu}A^{i}\right)\left( \partial_{\nu}A^{0} \right)
+ \left( 2 x_{9} + x_{15} + x_{16}\right)
\ \left(\partial_ {\mu}A^{0} \right)\left(\partial_ {\nu}A^{i}
\right)\\
& + \left(\partial^{0}A_ {\mu} \right)\bigl[2 \left( x_{9} + x_{15}\right)
\left(\partial^{i}A_ {\nu} \right)+
  \left( x_{7} + x_{14} +  x_{15}\right) \left(\partial_ {\nu}A^{i}
  \right)\bigr]\bigr\},
\end{split}
\end{equation}
whose vanishing leads to the solutions presented in Sec.~\ref{L7}.

\bibliographystyle{unsrt} 
\bibliography{bibli}

\end{document}